# Strategic Issues on Implementing a Software Process Improvement Program

### *Rogério Rossi & Kechi Hirama*
### *University of São Paulo, São Paulo, Brazil*

### rossirogerio@hotmail.com   kechi.hirama@usp.br

## Abstract

Software technology has high impact on the global economy as in many sectors of contemporary society. As a product enabling the most varied daily activities, the software product has to be produced reflecting high quality. Software quality is dependent on its development that is based in a large set of software development processes. However, the implementation and continuous improvement of software process aimed at software product should be carefully institutionalized by software development organizations such as software factories, testing factories, V&V organizations, among others. The institutionalization of programs such as a Software Process Improvement Program, or SPI Program, require a strategic planning, which is addressed in this article from the perspective of specific models and frameworks, as well as reflections based on software process engineering models and standards. In addition, a set of strategic drivers is proposed to assist the implementation of a Strategic Plan for a SPI Program which can be considered by the organizations before starting this kind of Program.

## Introduction

The software product is extremely relevant for modern society. It is currently an intrinsic product to an extensive number of daily activities, which does not allow software to be disregarded under any circumstances, on the finance, industrial, commercial, and educational sectors, among others.

Xu & Brinkkemper (2007) mention that the software is an everyday phenomenon and that the software product is able to substantially impact global economy. According to Pressman (2011), the feasibility of using software systems favors economy, being an important product as part of the delivery of other products and services.

In this sense, the software product is assumed to have its own production approaches that need to meet certain strategic objectives geared to the business of an organization. Slaughter, Levine, Ramesh, Pries-Heje & Baskerville (2006) consider that software development processes must meet certain organizational strategies.

The activities leading to the institutionalization of software process programs, i.e., the Software Process Improvement Program based on specific models or standards for software can also be strategically evaluated.

Some authors, such as Acuña & Juristo (2005), Zahran (1998) and Godbole (2005), present specific frameworks to institutionalize this kind of program to support actions aimed strategic planning of a Software Process Improvement Program within the organization.

Caputo (1998) mentioned that some potential strategic alignments can be adopted when considering maturity models for software, such as SW-CMM (Capability Maturity Model for Software). These alignments were directed exclusively to the CMM model and is currently greatly important that such strategic principles be considered in the adoption of other maturity models for software, such as current CMMI (Capability Maturity Model Integration), as well as to others models and standards that define practices for a Software Process Improvement Program such as MPS-BR (Brazilian Program for Software Process Improvement) or ISO / IEC 15504 standard.

However, as stated by Godbole (2005), programs of this nature, which seek to continuous transform the software process for its improvement, do not happen overnight. Such programs require investments, resources and clear expectations, so that they can generate the desired results.

The strategic aspects regarding strategic planning to enable a Software Process Improvement Program is the main goal of this research that also presents some frameworks and models for process improvement to software development organizations.

Software Processes Improvement must hence be included in the Strategic Planning through strategic actions aimed at obtaining a software product with better quality and offer better results to the organization and its customers. Within this context this article focuses on approaching strategic issues that promote and favor the implementation of a Software Process Improvement Program.

To meet its objectives, this article is organized as follows: section two presents the relation of the software product and competitive strategy; section three discusses the strategic management of software product by presenting specific frameworks viewing Strategic Planning for implementing a Software Process Improvement Program; section four lists the recognized software maturity models concerning strategic actions that can be identified within these models; section five presents a set of strategic drivers able to support the feasibility study of a Strategic Plan to institutionalize a Software Process Improvement Program;  section six presents a discussion and conclusion of the work, as well as suggestions for future work related to this research.

# The Software Technology as a Component for Competitive Strategy

The software product has specific characteristics and can be used in many sectors by contemporary society. Software automates the activities of the most diverse areas, such as commerce, industry, healthcare, education, entertainment, science, among many others, with the ability to streamline human actions and to favor the achievement of better results.

According Xu & Brinkkemper (2007, p. 4) "the software product is defined as a configuration of software components or software-based service that is released and traded in a specific market". Pressman (2011) consider that the software product is currently a single, unique product, which can be considered to manage the delivery of other products or services.

The software product is able to provide substantial increase in the overall economic capacity. There are several types of software product, such as commercial software, shelf software (COTS commercial off-the-shelf), software package, and open source software, among others, as presented by Xu & Brinkkemper (2007) and Pressman (2011). It is currently a product that enables high competitive ability for many organizations.

For Fitzpatrick (1996), to raise the software technology converting it into a product capable to more and better meet the needs of contemporary society, such a product and its components should be distinguished by high quality score. So, to improve software quality, the strategies adopted by an organization has to consider its process development.

Fitzpatrick (1996) consider the importance of strategies to improving software quality, either when it is produced to be used on a large scale as commercial software, or when it is produced for a specific customer, according to previous contractual agreements. He believes that the impact of quality should be considered on strategic planning for constructing and delivering better software product.

To achieve the strategic expectations, Paladini (2005) mention that the strategic actions correspond to those that have direct impact on the survival of organizations. For such strategic actions two dimensions can be considered: (1) spatial, which includes the organization as a whole and the environment in which it is embedded; and (2) temporal, where variables that change over time, such as technological progress, must be analyzed, as well the consumers' expectations, for example.

Thus, the spatial and temporal scope of the strategic actions should benefit the business strategies for the software industry at the moment of defining the coverage of the strategic planning: organizational, sectoral or functional.

According Slaughter et al. (2006), three levels of strategy can be distinguished in a company: organizational, by business unit, and functional; and these may influence the structures for selecting the best processes for developing software product and which, therefore, can significantly impact the software product.

Xu & Brinkkemper (2007) argue that from an organizational perspective, the software product should be considered according to the corporate strategies, product strategies and service strategies, as goals to reach the consumer market. Corporate strategies should be geared towards internationalization, investment, quality control and strategies for resource management. Strategies for product and service must support the corporate strategy with regard to the development and delivery of the software product.

As aforementioned, Slaughter et al. (2006) consider that strategies impacts the quality of the product at the organizational level, by business or functional unit, and Paladini (2005) mention that the construction of the strategic view of quality should consider some important points such as: 1) the construction of quality concept - consistently constructing this concept turning it into value for the entire organization; 2) the challenge quality means to face - competitive advantage in face of competitors; and 3) the contribution that quality aims to offer - operational contributions (increased productivity and decreased rework), tactics contributions (critical decision-making by a better prepared staff), and strategic contributions that not only seeks the survival of the company but also its continuous evolution.

These contributions transformed into actions at different organizational levels allowing competitive edge regarding the construction of the software product quality, be it produced on a commercial scale or developed on demand.

As mentioned by Slaughter et al. (2006), the strategic capacity should also be considered for selecting processes for developing software product. In this sense, the strategic development of the organization concerning practices of software process improvement is of great relevance and is explored in detail in the next section.

# Strategic Management for Software Process Improvement

A set of software engineering processes can greatly contribute to developer software organization resulting better software product. However, effective software process programs are sometimes started in developer software organizations only aiming to boost productivity.

These kinds of programs are composed by a variety of software development processes which is categorized by different approaches and applicable in different development phase within a software-making organization.

To Humphrey (1989), Slaughter et al. (2006), and Zahran (1998) the software development processes correspond to a set of activities, methods and practices used to develop software. Acuña & Juristo (2005) believe that software processes involve a series of steps, methods, procedures, techniques and tools used to develop, deliver and maintain software. IEEE (1990) define the software development process as the translation of user needs into software requirements, transforming these into design, implementing the design in code, testing the code, and sometimes installing and checking out the software operation. Humphrey (1989) believe that software processes have a strong dependence on human factors, especially knowledge, skills and people's attitudes.

For Acuña & Juristo (2005), many organizations consider and possibly implement Software Process Improvement Programs in order to obtain benefits that turn to business as increasing quality, lowering costs by improvements in productivity, reducing development cycle time, and improving customer satisfaction.

Humphrey (1989) consider that improving the software process corresponds to a migration of the current state to a desired state, refining the methods, procedures and tools to improve the software process.

However, Software Process Improvement Programs need structured approaches to be implemented and maintained. Acuña & Juristo (2005) warn that the literature is scarce when seeking ways to manage organizational change to Software Process Improvement Program.

Schulmeyer & McManus (1999) present a relevant relation between Software Process Improvement concepts and the Software Engineering Institute (SEI). Such concepts have been widely studied by the SEI, and detailed actions related to these concepts allowed the SEI to disclose maturity models for software product that culminated in the current CMMI-DEV (Capability Maturity Model Integration for Software Development) as verified in SEI (2010).

However, the implementation of models aimed at software process improvement, such as CMMI, sometimes requires conditions and strategic capability of the organization, as can be verified in differentiated frameworks proposed by Acuña & Juristo (2005), Zahran (1998) and Godbole (2005) and presented in Table 1.

Zahran (1998) deems that an appropriate environment must be considered to implement and to maintain components and mechanisms aimed at continuous software process improvement. These mechanisms should promote the establishment of organizational culture dedicated to managing processes and infrastructure to support the development of projects within the organization.

**Table 1: Frameworks with strategic proposals for a Software Process Improvement Program**

| Author | Proposed Framework |
|---|---|
| Zahran (1998) | **Framework for Software Process Improvement:** <br> 1) Infrastructure for Software Process Improvement <br> 2) Roadmap for Software Process Improvement <br> 3) Evaluation methods for software process <br> 4) Plan for Software Process Improvement |
| Acuña & Juristo (2005) | **Pragmatic Model for Implementing a Program for Software Process Improvement** <br> 1) Preparation <br> 2) Planning <br> 3) Implementation <br> 4) Institucionalization |
| Godbole (2005) | **Strategic Drivers for Implementing a Software Process Improvement Program** <br> 1) Developing proposals for process improvement <br> 2) Evaluating developed proposals <br> 3) Defining project for the program <br> 4) Defining a schedule for implementing defined proposals <br> 5) Obtaining management commitment |

To create an effective culture of software processes and infrastructure to manage the Software Process Improvement Program, Zahran (1998) present a framework for Software Process Improvement (Table 1) detailed as follows:

1. **Infrastructure for Software Process Improvement**: basically considering two infrastructures to support the software process; the first one organizational and managerial, the second, technical;

2. **Roadmap for Software Process Improvement**: a logical approach with identified steps should be defined to provide effective implementation of software processes. In this case, models and standards can be adopted as the CMMI or ISO / IEC 15.504 or adopting customized versions to meet the needs of the organization;

3. **Evaluation Methods for Software Process**: this evaluation method must be applied to assess the current situation of software processes used by the organization. Perhaps this assessment is performed according to what is prescribed in the roadmap adopted (item 2). Improvement actions should gradually satisfy what is established by the roadmap (model and / or standard adopted by the organization);

4. **Plan for Software Process Improvement**: this plan involves the transformations identified during the evaluation phase (item 3). Actions must be understood and specified so that they are effectively applied to the processes to effectively achieve the desired improvements.

Acuña & Juristo (2005) propose a model for Managing Organizational Change for Software Process Improvement. The model refers to a pragmatic approach to organizational change initiative to implement a Software Process Improvement Program presented in Table 1 and detailed as follows.

1. **Preparation**: an insight into the desired changes must be developed and guided by the processes that need improvement, why they need improvement, as well as defining a priority order for completing the improvements. Typically, a formal evaluation is conducted to determine areas needing improvement;

2. **Planning**: includes formulating measurable objectives to achieve the improvements in software processes aligned with the view of change previously determined and the projection of the efforts to be undertaken to achieve such improvements. This involves defining manageable activities, respecting time constraints and the definition of roles and responsibilities to make the changes suggested in the Preparation phase;

3. **Implementation**: this phase should operationalize the plan and achieve the goals of the software process improvement as defined in the planning time. According to the priority given to items, at this phase, the planned improvements should be implemented using the resources provided and should communicate the benefits of the improvements made to perpetuate the view of continuous process improvement;

4. **Institutionalization**: this phase should properly verify that the change (improvement) is permanent in the organization. This is ensuring that the process has been definitely improved, used by all employees of the organization, except for pilot projects, so that the benefits that were intended for the business are achieved.

A Software Process Improvement Program should be aligned with the organizational business goals to be successfully implemented. The Process Improvement Plan must be developed after conduction of a self-assessment to determine the specific skills to perform the necessary improvements.

In this sense, Strategic Drivers for Implementing a Software Process Improvement Program are proposed by Godbole (2005) as another framework that favors the implementation of this kind of programs, as shown in Table 1. For Godbole (2005) the strategic actions for the implementation of such a program should consider the following steps:

1. Developing proposals for process improvement;
2. Evaluating developed proposals;
3. Defining Project participants;
4. Defining a schedule for implementing the proposals; and
5. Obtaining management commitment.

Strategic drivers mentioned by Godbole (2005) should be considered in the Strategic Planning phase for implementing the Software Process Improvement Program and attention should be given to the following actions that favor the success of the program:

1. Strong relation with the organization's business processes;
2. Selection and involvement of the right people to implement this plan;
3. Productive communication and sharing of ideas among those involved in the program; and
4. Approach that provides an objective view on the proposed actions.

These actions bind the Software Process Improvement Program to other processes of the organization's business involving the necessary teams for conducting improvements, considering effective communication mechanisms to disseminate the actions towards improvement and the advances achieved.

# Strategic Perspectives According to Software Maturity Models

The initial actions to implement a Software Process Improvement Program vary according to the organization, with its organizational objectives, resources, partners, clients, with budgets and deadlines, etc. The conditions of these aspiring organizations sometimes require careful evaluations for adopting models and standards to be used in their structures for developing such a complex product as the software product.

Thus, analysis at different levels and with different objectives should be undertaken to facilitate the implementation of such a program, especially strategic analyses that address the viability of these programs.

By using models and standards able to facilitate the implementation of these kinds of programs, the organizations do not necessarily changing elements regarding strategic planning to implement these programs. Likewise, these models and standards do not consider practices explicitly concerning strategic planning.

There are many models and standards to support the organization in these types of corporate actions to implement a Software Process Improvement Program, e.g.: 1) CMMI (Capability Maturity Model Integration) as SEI (2010 ), 2) MPS-BR (Brazilian Program for the Software Process Improvement) that can be verified in SOFTEX (2012), 3) ISO / IEC 12207 that establishes a common framework for processes of software life cycle (ISO, 1995), and 4) the BOOTSTRAP model, similar to the SW-CMM model characteristics, among others.

Considering these examples, their structures allow observing practices and methods that must be implemented according to the principles of process engineering, i.e., each of them strongly aimed the implementation of continuous process improvement programs to the software product in a process management approach.

The models and standards shown above as examples also provide a set of specific practices that are structured and should be institutionalized to promote a continuous software process improvement and maturity.

Aiming to detail the strategic issues provided by these models and standards, the following structures (CMMI, MPS-BR, and ISO 12207) will be presented in more detail to emphasize the practices present in these models, concerning the issues for institutionalizing a Software Process Improvement Program.

### CMMI (Capability Maturity Model Integration)

The CMMI (Capability Maturity Model Integration), as shown in SEI (2010) and (Chrissis, Konrad & Shrum, 2004) is a complete framework for the software product and process that has some maturity models in its content. From these models, the model to be highlighted in this article is the CMMI-DEV (CMMI for Development), because it presents the specific practices aimed at the development process of the software product.

The CMMI-DEV in its 1.3 version (SEI, 2010) consider the processes divided into four groups: 1) Process Management, 2) Project Management 3) Product Engineering, and 4) Support. Thus, in this work, the Process Management group is considered for analysis of the practices linked to strategic issues.

Practices associated with the processes of this process group can offer some relevant resources to the organization that seeks to implement a Software Process Improvement Program as evidenced in (SEI, 2010). They describe structural elements for process engineering. The practices are capable of supporting the organization and the organizational culture of management by processes, but make no explicit reference to analysis or strategic actions related to the implementation of the Software Process Improvement Program.

Chrissis et al. (2004) present a framework for Software Process Improvement which regards a Software Strategic Planning Group associated with a Strategic Plan at corporate level as a major component of this structure. According to Chrissis et al. (2004), this structure must be able to support the implementation of the Software Process Improvement Program to be institutionalized. However, this proposal structure is simply considered a good practice, sometimes being considered in process improvement programs implemented in North American corporations.

Caputo (1998) describe that the cycle presented below must be followed to implement a Software Process Improvement Program (considering the extinct SW-CMM model) to cause the constant improvement of processes:

- Forecasting - actions needed to verify what has been done and what should be done;
- Coding - actions regarding the decisions about what should be done and the formalization related to how to do it;
- Publicizing - making changes and improving performance.

In this cycle, these actions favor the simplified strategic movement to start the Software Process Improvement Program. Although Caputo's (1998) proposal has been performed based on the SW-CMM model, it can be applied to other maturity models for software processes, such as CMMI.

## MR-MPS (Reference Model of Brazilian Process Improvement Program for Software)

The MPS-BR Program considers a range of models and guides in its structure. According to SOFTEX (2012), the program considers the Reference Model, the Business Model and the Evaluation Method. Specifically considering the Reference Model (MR-MPS) and its Software Evaluation Guides for each maturity level (levels G to A); it appears that it presents a number of practices that must be met favoring a Software Process Improvement Program.

Such practices are divided into seven maturity levels, from the lowest level (level G) to the highest level (level A). Its specific processes are described from level G to level B, given that level A does not present specific processes, as shown in Table 2.

**Table 2: MPSBR – Processes defined by Maturity Levels**

| Level | Related process |
|---|---|
| A | (no new processes are added) |
| B | Project Management (new outcomes) |
| C | Decision Management<br>Risk Management<br>Development for Reuse |
| D | Requirements Development<br>Product Design and Construction<br>Product Integration<br>Verification<br>Validation |
| E | Human Resources Management<br>Process Establishment<br>Process Assessment and Improvement<br>Project Management (new outcomes)<br>Reuse Management |
| F | Measurement<br>Configuration Management<br>Acquisition<br>Quality Assurance<br>Project Portfolio Management |
| G | Requirements Management<br>Project Management |

During the evaluation of maturity levels, the processes are assessed according to process attributes. These vary for each level; fewer process attributes are checked at the lowest level but increase as higher level of maturity is sought.

Hence, an analysis as to the process and the process attributes that favored this research related to strategic assessment the program provides for organizations adopting this model as verified in (SOFTEX, 2012). In this sense, there are no specific practices within the model that specifically concern the strategic dimension.

The model requires that one specific practice is implemented from level E (GPR practice 22) and that a process is defined to manage the strategy of adapting the processes defined by the organization. Furthermore, in one of its attributes for measuring the processes, that it considers the need to check the implementation strategy for the improvements planned for the organization. However, these practices do not favor the strategic assessment related to the implementation of a Software Process Improvement Program *a priori*; it is a practice that treats strategic issues when the organization is already involved in a Process Improvement Program and this program is already being conducted by the organization.

### ISO/IEC 12207

The ISO/IEC 12207 standard establishes a common framework for processes of software life cycle (ISO, 2004), that can be referenced by diverse software organizations (providers, consumers and developers). It provides a set of processes that can be used for define, control and improve processes of life cycle.

It was published, initially, in August, 1995 (ISO, 1995) and updated with Amendment 1 in 2002, compatible with ISO/IEC 15504, ISO/IEC 14598 and ISO/IEC 15939 (ISO, 2002). Further, it establishes a Process Reference Model in accordance with the requirements of ISO/IEC 15504-2 standard. The Amendment 2 resolves some technical defects and editorial issues in Amendment 1 (ISO, 2004).

The ISO/IEC 12207:2008 standard applies to the acquisition of systems and software products and services and may be used stand alone or jointly with ISO/IEC 15288. It supplies a process reference model that supports process capability assessment in accordance with ISO/IEC 15504-2 (ISO, 2008).

# Strategic Drivers for Software Process Improvement

Although several frameworks have been created to contribute to the Software Process Improvement Program, it is feasible that strategic actions are considered as a goal for institutionalizing such programs.

Garvin (1992) propose for quality eras: Inspection, Quality Control, Quality Assurance and Strategic Quality Management, which allows organizations of many sectors to evolve their approach of quality management from inspection actions to strategic quality management.

For the software industry, such principles benefit the approach related to a Software Process Improvement Program. Some authors such as Godbole (2005), Zahran (1998) and Acuña & Juristo (2005) propose structures (as evidenced in Table 1) that favor strategic actions to verify the feasibility and objectives as well as to direct activities aimed the implementation of a Software Process Improvement Program.

As aforementioned, such structures have some common points and one of them concerns to the necessities of implementing a plan for this type of program. Zahran (1998) consider that planning would facilitate the implementation of the program. Acuña & Juristo (2005) consider that measurable goals should be defined in the adoption of a program like this. For Godbole (2005), planning actions should be considered in a previous phase to the implementation of these programs to ensure that the proposed improvements expected are reached.

Thus, it is possible to consider that the implementation of a Software Process Improvement Program must be done with strategic reasoning that allows assessing the viability of the program, defining the program goals, resources, investments, expected results, etc.

Godbole (2005) mention that a program of this type, i.e., the transformation of organizational processes, to be improved, adequate and implemented in order to generate the best results, is not instantaneous. Such actions require commitment, time and organizational effort to create the necessary change.

The strategic actions that should be directed to favor the implementation of a Software Process Improvement Program can be based on a set of strategic drivers for software quality management called SPSPI (Strategic Planning for Software Process Improvement) to collaborate to the strategic planning of a Software Process Improvement Program.

In order to provide actions towards a strategic evaluation of quality for a software industry, SPSPI presents some drivers to support the strategic planning for software quality aiming at supporting organizations to institutionalize a Software Process Improvement Program. The group of drivers associated with SPSPI is presented below and detailed as follows:

- Conceptualization of Quality concept for the organization;

- Definition and proposition of Quality Plan;

- Alignment of Quality Plan with business goals;

- Definition of expectations and results;

- Definition of the approach to Software Process Improvement Program;

- Selection, analysis and adoption of models and standards to support the program;

- Acquisition of commitment from people/staff;

- Definition of measurement and indicators;

- Allocation of financial resources and investment.

Although these drivers establish strategic actions for a Software Process Improvement Program, the strategic evaluation and its outcome should culminate in a Strategic Quality Plan, and be linked to other organizational strategic plans, sometimes being an integral part thereof.

The strategic drivers must cooperate to the strategic view of the organization and, in this sense; the quality conceptualization to the organization should be a major factor. What the organization means by quality, how it will seek to address this issue, how this characteristic must be present in the results must be cleared during this strategic analysis. The organization should clearly establish the concept so that it can foster further action; they may hence consider classical concepts about software quality. Paladini (2005) consider that the quality concept should be formulated by the organization so that it can add some value to the results. Relevant links to this conceptualization are mechanisms for measuring and reporting the results derived from such measurements.

Planning the quality to delineate the Software Process Improvement Program to be institutionalized involves defining objectives, resources, people, tools, models, etc., that contribute to the program implementation. This planning should be consistent with other business objectives, i.e., other organizational goals in a corporate view. The program seeking the improvement of software processes, which affects the software product, certainly has significant organizational impact; this should be considered during the strategic assessment of how these will be integrated towards better results.

The strategic planning of a program of this magnitude should thus set the expectations and results for the short, medium and long term. What outcomes are to be achieved, what indicators should be created or made possible to provide objective measurement mechanisms to present the results previously pursued.

The approach of continuous process improvement itself must be carefully analyzed and is linked to another driver, which concerns the viability of adopting models and standards that support actions to improve processes. In this case, a rigorous study justifying the actions of the organization should be undertaken not laying the responsibility for the program on a given model, standard or norm that supports programs of this nature. The process improvement is independent of models and standards; yet, they are able to favor the program implementation. Therefore, the organization should have specific goals to improve their processes. If the approach is bound to adopt specific models or standards, the organization must define which model and/or standard is more feasible for its Software Process Improvement Program considering, for example, the necessary investments, the market share to achieve, the resources to be used, etc.

Some questions should be asked to allow real change and process improvement. If these activities are guided by the adoption of specific models and standards, this adoption must address the objectives of the organization. The program's specific goals and corporate objectives should be considered, given the investment and the allocation of financial resources to be performed and the perspectives regarding the changes and improvements that should be reached.

## Discussion and Conclusion

Strategic evaluation accounts for the successful programs and projects implementation in many sectors. This type of evaluation and actions for strategic planning are considerably relevant for the Software Process Improvement Program.

Strategic actions through specific drivers are able to provide planning and to support the institutionalization of a Software Process Improvement Program. Based on a specific model like CMMI, or simply performed without being based on a model, strategic planning should be conducted by organizations that seek to institutionalize a Software Process Improvement Program.

As verified from the frameworks presented by expert researchers in this knowledge area, the authors proposed structures which can contribute to strategic thinking aimed at Software Process Improvement programs. However, from the analysis aimed specifically at continuous process improvement models for software, there are no specific requirements identified by these models related to strategic actions regarding the institutionalization of a Software Process Improvement Program, which allows concluding that the strategic evaluation and planning are not really the purpose of these types of models.

According to the analysis and conducted research, it is possible to propose a set of strategic drivers (SPSPI) to be used by a software development organization to strategically support the previous evaluation of the implementation of a Software Process Improvement Program. The strategic drivers can collaborate with the organizations in defining their strategic plan for institutionalizing these types of programs.

The use of this set of drivers can be performed by companies of all sizes and sectors; however, reflection on future researches on this subject suggest that: 1) a survey to identify companies that already have an institutionalized Software Process Improvement Program in their environment to improve software processes, investigating whether a strategic plan was considered for the institutionalization, and at what level this planning was done; and 2) a survey to identify companies that do not already have such programs, if they would consider conducting a strategic planning based on the proposed strategic drivers, or even whether they would consider using some of the frame-works presented in Table 1, viewing the strategic movement for implementing a Software Process Improvement Program.

# Biographies

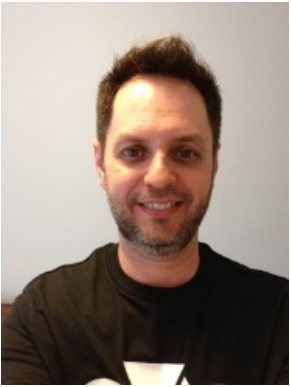

**Rogério Rossi** received his B. S. in Mathematics by the University Center Foundation Santo André as he also has a M.S. and Ph.D. in Electrical Engineering, both by Mackenzie Presbyterian University. He is in a Postdoctoral Program at the University of São Paulo developing researches that are related to Complex Systems, Big Data and the Internet of Things (IoT).

He is an Adjunct Professor for Information Technology and Computer Science courses of graduate and undergraduate programs in São Paulo. He has done research on the fields of software quality, and quality for digital educational solutions and he also has some publications on this area.

He is a member of IACSIT (International Association of Computer Science and Information Technology) and he worked as a reviewer for InSite Conferences'2013 and 2015, and e-Skills Conference'2014; as he also presented his papers in the InSite Conferences in Montreal, Canada (2012) and Porto, Portugal (2013).

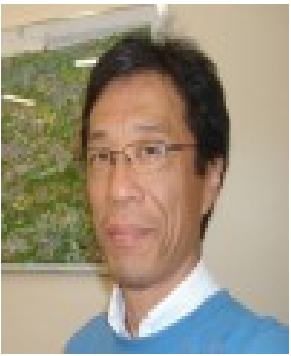

**Kechi Hirama** received his B.S., M.S., Ph.D. and Associate Professor degrees in Computer Engineering from Escola Politécnica of the University of São Paulo, São Paulo, Brazil in 1980, 1988, 1996 and 2008, respectively.

He worked 15 years in the Control and Automation area in research organizations and since 1996 he has been a Professor of the Department of Computer and Digital Systems Engineering of Escola Politécnica of the University of São Paulo.

His interests include Complex System, System Dynamics, Big Data and Internet of Things (IoT).